\documentstyle[preprint,aps]{revtex}
\begin{document}
\tightenlines

\title{Pion-Baryon Couplings}

\newcommand{\xbf}[1]{\mbox{\boldmath $ #1 $}}

\author{Alfons J. Buchmann$^a$\thanks{email:alfons.buchmann@uni-tuebingen.de} 
and 
Ernest M. Henley$^b$\thanks{email:HENLEY@nucthy.phys.washington.edu}
\\
\vspace {0.8 cm}
$^a $ Institute for Theoretical Physics, University of T\"ubingen, \\
D-72076 T\"ubingen, Germany\\
and \\
$^b$ Department of Physics and Institute for Nuclear Theory, \\
University of Washington,
Box 351560, Seattle, WA 98195, USA}

\maketitle

\begin {abstract}

We have extended and applied a general QCD parameterization method
to the emission of pions from baryons. 
We use it to calculate the strength and sign of  the coupling of pions 
to the octet and decuplet of baryons. Certain relations between octet
and decuplet couplings are pointed out.

\end{abstract}
\newpage

In 1989 Morpurgo \cite{Mor89,Mor92} 
introduced a parameterization for the properties of
hadrons, which expresses masses, magnetic moments, transition amplitudes,
and other properties of the baryon octet and decuplet in terms of a few 
parameters. The method uses only 
general features of QCD and baryon descriptions in terms of quarks. 
Recently, Dillon and Morpurgo have shown that the method is independent
of the choice of the quark mass 
renormalization point in the QCD Lagrangian \cite{Mor92}.   
They have also extended the method to nucleon electromagnetic form factors
and radii \cite{Dil99}.

In addition to the electromagnetic properties of baryons,
i.e., the interaction with an external photon field,
it is possible to consider the interaction of a baryon with an external 
pion field, and to calculate the pion-baryon couplings. 
Due to the internal quark structure of the pion, this problem
is rather different from the ones treated in Refs. 
\cite{Mor89,Mor92,Dil99}. Despite the additional
difficulties due to the pion's size and mass, the Morpurgo method
is nevertheless applicable here, as we argue below.

The Morpurgo method is based on the following considerations.
For the observable at hand one formally writes a QCD operator 
$\Omega$ and QCD eigenstates expressed explicitly
in terms of quarks and gluons. This matrix element
can, with the help of the unitary operator $V$, be reduced to a calculation
in the basis of auxiliary (model) three-quark states $\Phi_B$ 
\begin{equation}
\label{map}
\left \langle B \vert \Omega \vert B \right \rangle = 
\left \langle \Phi_B \vert
V^{\dagger}\Omega V \vert \Phi_B \right \rangle =
\left \langle W_B \vert
{\cal O} \vert W_B \right \rangle \, .
\end{equation}
Both the unitary operator $V$ and the model states $\Phi_B$ are defined in  
Ref.\cite{Mor89}. 
The $\Phi_B$  
are pure $L=0$ three-quark states excluding any quark-antiquark
or gluon components. $W_B$ stands for the standard three-quark
$SU(6)$ spin-flavor wave functions.
The operator $V$ dresses the auxiliary states
$\Phi_B$ 
with $q\bar q$ components and gluons and thereby generates
the exact QCD eigenstate. Furthermore, the operator $V$ 
contains a Foldy-Wouthuysen transformation, which transforms 
the original 4-component Dirac spinor $B$  into a two-component 
Pauli spinor contained in $W_B$. 

One then writes the most general expression for ${\cal O}$ 
compatible with the space-time and inner QCD symmetries.
The orbital and color space matrix elements are absorbed 
in unknown parameters multiplying the various invariants
appearing in the expansion of ${\cal O}$. 
As an example, for the squared charge 
radius ${\cal O}$ we need a scalar 
operator linear in the quark charge  $Q_i$ \cite{Dil99}. 
The lowest order one is just $\sum_i Q_i$. The
next higher order expression is $\sum_{i \neq j} Q_i {\xbf{\sigma}}_i \cdot 
{\xbf{\sigma}}_j$, where the sum is over all quarks.
A three-body term is $ \sum_{i \neq j \neq k} 
Q_i {\xbf{\sigma}}_j \cdot {\xbf{\sigma}}_k$, so that the full
expression reads
$$ r^2_B = A \sum_i Q_i + B \sum_{i\ne k} Q_i \xbf{\sigma}_i \cdot\xbf{\sigma}_k
+C \sum_{i\ne j \ne k} Q_i \xbf{\sigma}_j \cdot \xbf{\sigma}_k + \cdots \, .
$$
The expectation value of the three-quark term is expected to be  $\leq 1/3$ 
of the two-quark term, which in turn should be $\leq 1/3$ of the one-body
term. The reasons for this hierachy of expressions are discussed in Ref. 
\cite{Mor92}.

Coming back to the nucleon pion coupling and writing the standard
effective pseudovector coupling 
\begin{equation}
H = -\frac{f}{m_{\pi}} \;\bar{\psi}\, \gamma_\mu \gamma_5\, \psi 
\partial ^\mu \,\vec{\phi}\cdot \vec {\tau}   \, ,
\end{equation}
it appears that,
for a nucleon at rest and in the limit of small 
four-momentum transfer to the pion, the operator $\Omega$ --no matter 
how complicated-- must be such that
\begin{equation}
\label{pionquark}
\left \langle N \vert \Omega \vert N \right \rangle = {f \over m_{\pi}}
\xbf{\sigma} \cdot \xbf{\nabla}  \vec{\tau} \cdot \vec{\phi} \, . 
\end{equation}
Eq.(\ref{pionquark}) defines the coupling constant $f$ of the (point) 
pion field $\vec{\phi}$ to the nucleon, $\xbf{\sigma}$ is the nucleon spin
and $\vec{\tau}$ the isospin matrix. It is understood that the right-hand
side of Eq.(\ref{pionquark}) is calculated between the spin-isospin state
of the nucleon. 
In the limit of small four-momentum
transfer $\xbf{\nabla}\cdot \vec{\phi}=i {\bf k} \cdot \vec{\phi}$.
As noted in Ref.\cite{Mor89}, even if the right-hand side is non-covariant,
referring to the rest frame, the theory is relativistically
complete. There can be no other spin-structure for
the pion-nucleon interaction, in the limit of small ${\bf k}$.

Because the right-hand side of Eq.(\ref{pionquark}) has a structure 
similar to the magnetic moment operator,
we proceed as in Refs. \cite{Mor89,Dil99}. 
As shown in detail by Morpurgo \cite{Mor89}, the most general 
parameterization of the axial vector coupling operator 
${\cal O}$  for octet and decuplet baryons Eq.(\ref{pionquark}) 
can be classified in terms of
one-quark, two-quark, and three-quark terms plus Trace (closed loop) terms. 
The latter are not present here.

For the case under discussion the one-body operator can be taken as
\begin {equation}
{\mathcal{O}}_1 =A_1\sum_i {\vec \tau^i} {\xbf{\sigma}}^i
\cdot {\bf k}\;,
\end {equation} 
where the sum is over the 3 quarks present in the auxiliary state 
$\Phi_B$. For our 
purpose, we can use
\begin{equation}
{\mathcal{O}}_1 = A_1 \sum_i \tau_3^i  \sigma_z^i k_z \;.
\end{equation} 
With this ${\mathcal{O}}_1$, the most general two-body term is given by 
\begin {equation}
{\mathcal{O}}_2=A_2\sum_{i, j\neq i} \tau_3^i \sigma_z^j k_z \;.
\end{equation} 
Although one can make up further two-body terms, e.g., $(\vec{\tau^i}  \times
\vec{\tau^j})_3 \;({\xbf{\sigma}}^i  \times {\xbf{\sigma}}^j)_z k_z$, 
Dillon and Morpurgo \cite {Dil99} have shown that they are not
independent. One can also make up three-body terms as 
${\cal O}_3=A_3\, \tau_3^i\, \sigma_z^i\, k_z\ 
{\xbf{\sigma}}^j \cdot {\xbf{\sigma}}^k$ and others; 
we shall neglect them here. 
In that case, the operator we need is 
\begin {equation}
\label{quarkop}
{\mathcal{O}}={\mathcal{O}}_1 + {\mathcal{O}}_2 \; .
\end{equation} 

The auxiliary (model) states $\Phi_B$ used for calculating the expectation 
values of the operator ${\cal O}$ coincide 
with the familar $SU(6)$ states multiplied    
by a spatial wave function with orbital angular momentum $L=0$.
The $SU(6)$ states are listed, e.g., in Ref.\cite {Clo}. 
We only need the completely symmetric spin-isospin part, $W_B$. 
The radial and color parts of the matrix elements are absorbed in 
the constants $A_1$ and $A_2$.

We present, separately, the quark model matrix elements of the operator 
in Eq.(\ref{quarkop}) to first order (one-body terms) and second order 
corrections (two-body terms) in Table I.
$\,A_1$ and $A_2$ are constants to be determined below, 
and $r$ is the ratio $\frac {m_u + m_d}{2 m_s}= \frac{m_u}{m_s}$, 
where $m _i$ is the mass of quark i. The reason we include $r$ 
in the two-body term is
that a two-body gluon exchange between particles 1 and 2 would be 
inversely proportional to the masses of the two quarks. 
This approximate way to take into account SU(3) symmetry breaking 
works quite well for magnetic moments \cite{Wag95}.  
We neglect SU(3) symmetry breaking in the one-body term.
A more rigorous treatment of SU(3) symmetry breaking requires
the introduction of two constants \cite{Mor89}.

In order to obtain from the
quark model matrix elements in Table I the conventional pion-baryon 
couplings, additional overall factors are needed. 
The pion-octet baryon couplings are defined for spin projection 
$m_s=+1/2$ and maximal isospin projection of the corresponding 
baryon level operator
evaluated between baryon spin and isospin wave functions.
Similarly, 
the $\Delta\Delta\pi$ coupling is defined for maximal spin and 
isospin projection \cite{Bro75}. 

If we neglect the two-body operators, then the pion 
coupling to the nucleon is sufficient to fix the unknown constant $A_1$ 
of the  theory. When  two-body corrections are included, we use the decay 
of the $\Delta (1232)$ to fix the second constant $A_2$. 
The $N\Delta\pi$ coupling is of the form
\begin{equation}
H_{N\Delta\pi} 
= - \frac{f_{N\Delta\pi}}{m_\pi}\, 
\mathbf{ S^{\dagger} \cdot {\bf{\nabla}} \vec{T}^{\dagger}} 
\cdot \vec{\phi} + h.c. \; ,
\end{equation}
where ${\bf S}^{\dagger}$ and $ \vec{{\bf T}^{\dagger}}$ are 
transition spin and isospin operators; they 
are defined such that their matrix elements are simple Clebsch-Gordan
coefficients \cite {Eri}. The coupling $f_{N\Delta\pi}$ is taken to be 
2 f, which gives the $\Delta(1232)$ its experimental 
width of about 130 MeV \cite{Eri}. 

Without two-body terms, $A_1$ is 
fixed by $\frac {f^2}{4 \pi} =0.08$, and $A_1 =(3/5) f$.  The 
first entry in Table I is equal to $g_A$, the 
axial coupling constant in the additive quark model. In this 
approximation, one obtains the well known additive quark model result 
for the $N\Delta\pi$ coupling $f^2_{N\Delta\pi} = (72/25) f^2$.
If we include the two-body terms, then we also need the empirical
relation $f_{N\Delta\pi} \approx 2 f$ to fix $A_2$. In this case, we obtain 
$A_1 \approx 0.53 f$ and $A_2 \approx -0.18 f$, 
so that $A_2/A_1 \approx 1/3$, a quite substantial correction to the
additive quark model. For exact SU(3) symmetry, $r =1$, 
but if SU(3) is broken, then $r \approx 330/550 \approx 0.6 $.

Table II lists the various couplings in terms of $\,f\,$, the $\pi^0 p$ 
coupling constant, to first order and to second order with and without the 
inclusion of r.  We recall that for the decuplet-octet transition 
couplings, baryon level spin and isospin Clebsch-Gordan coefficients 
$(1\, 0 \  S \, S_z \vert S' \,S_z')\,\, (1\, 0 \  T\, T_z \vert T'\, T_z')$ 
are needed in order to convert the quark level matrix elements in Table I 
to the (baryon level) coupling constants listed in Table II.
Here, $S(T)$ refers to octet, and $S'(T')$ to decuplet baryons.
Similarly, in order to obtain the baryon level decuplet-decuplet couplings
one uses a conversion factor 
$1/(T_z'\, S_z')$ \cite{Bro75}.
For example, the entry for $\Delta^{+}$ in Table II contains a 
factor $4/3$ to go from quarks to the $\Delta^{+}$. 
Without SU(3) breaking the decuplet-decuplet $(DD)$ couplings can then
be generically written as $f_{DD}=4/3(A+2B)$, which implies, e.g.,
$f_{\Delta^{++}\Delta^{++}\pi^0}=f_{\Delta^{+}\Delta^{+}\pi^0}$.

Our results satisfy the following relation in the SU(3) symmetric case
\begin{eqnarray}
\label{rel1}
f_{\pi^0 p}+ f_{\pi^0 \Xi^0} & = &  f_{\pi^0 \Sigma^+}\; .
\end{eqnarray}
Note that the second-order correction is especially
large for the $\Xi^0$, because the coefficient in front of $A_2$ is 4 times 
as large as that in front of $A_1$. 
The two-quark operator ${\cal O}_2$ changes the 
$\pi^0 \Delta^{+} \Delta^{+}$ coupling from the first order
value $f_{\pi\Delta\Delta}=(4/5)f$
to the total result $f_{\pi\Delta\Delta}=0.23 f$. 
We agree with Brown and Weise \cite{Bro75} for those pion couplings they 
calculated using a one-quark operator, namely to the nucleon and $\Delta$. 

The $\pi \Sigma \Lambda$ and the 
$\pi \Sigma^* \Lambda$ couplings remain unaffected by SU(3) 
symmetry breaking. Irrespective of the value of $r$ our 
octet-decuplet transition couplings satisfy the sum rule
\begin{equation} 
\label{rel2}
\sqrt{2} f_{\Delta^+ p}= f_{\Xi^{*0}\Xi^0} + 
\sqrt{6} f_{\Sigma^{*+}\Sigma^+} - f_{\Sigma^{* 0}\Lambda^0}.
\end{equation}
This relation is not new. It has been 
derived before \cite{Bec64} using SU(3) symmetry and its
breaking to first order. 

Next, we compare our results for the transition couplings 
to those obtained in the large 
$N_c$ approach \cite{Das94}. Because the couplings in \cite{Das94}  
are computed for $\pi^+$ emission, we calculate the  
matrix elements in Eq.(\ref{map}) with $\tau_{z}$ 
in Eq.(\ref{quarkop}) replaced by $\tau_{+}$. We then obtain for 
the transition $\Delta^{++} \to p \pi^+: 4\sqrt{3}(A-B)/3$,
for $\Sigma^{* +} \to \Sigma^0 \pi^+:-2\sqrt{2}(A-(2r-1)\,B)/3$, 
and finally for $\Sigma^{* +} \to \Lambda^0 \pi^+:-2\sqrt{6}(A-B)/3$.   
By taking ratios of two transition couplings we get for the case $r=1$
\begin{equation}
{\Delta^{++} \to p  \over \Sigma^{* +} \to \Sigma^0 } =   
-\sqrt{6} \ \ \ \ (-3.06)  \; , \ \ \  \ \  \qquad
{\Sigma^{* +} \to \Sigma^0  \over \Sigma^{* +} \to \Lambda^0 } =   
-{1 \over \sqrt{3}} \ \ \ \ (-0.46) \; 
\end{equation}
The numbers in parentheses include SU(3) symmetry 
breaking in the two-quark term $(r=0.6)$. 
These results are in agreement with those obtained in the 
large $N_c$ approach \cite{Das94}, including the next-to-leading order 
corrections, which is undoubtedly more than a numerical coincidence.

In Table III we compare the couplings we obtain with the 
inclusion of the two-body terms to those derived by other means.
The values of the coupling constants from Stoks and Rijken (SR) 
\cite {Sto} are obtained from fits to baryon-nucleon scattering  data and 
one-boson exchange potentials. The difference in sign from SR in the entry 
to Table III is because they use the coupling for 
$\pi^+\,\Sigma^+\,\Lambda^0$. The SR couplings are essentially the same as 
those of Maessen, Rijken, and de Swart \cite{Mae}. The columns labeled 
KDOL \cite{Kim} are obtained from QCD sum rules with the use of SU(3) and 
``beyond'' SU(3) by correcting for mass effects. Our values tend to be closer 
to those of Stoks and Rijken \cite{Sto}. The latter also satisfy 
Eq.(\ref{rel1}) in contrast to the SU(3) symmetric values of Ref.\cite{Kim}.

Finally, we point out certain analytical relations between 
octet and decuplet baryon couplings to pions that emerge from our theory
(neglecting three-quark terms) 
\begin{eqnarray}
\label{rel3}
f_{\pi^0 p}- {1\over 4} f_{\pi^0 \Delta^+\Delta^+} & = & 
{\sqrt{2} \over 3} f_{\pi^0 p \Delta^+} \nonumber \\ 
f_{\pi^0 \Sigma^+}- {1\over 2} f_{\pi^0 \Sigma^{* +}\Sigma^{* +}} & = & 
{1 \over \sqrt{6}} f_{\pi^0 \Sigma^{* +} \Sigma^+} \nonumber \\
f_{\pi^0 \Xi^0}- {1\over 4} f_{\pi^0 \Xi^{* 0}\Xi^{* 0}} & = & 
{ 1 \over 3} f_{\pi^0 \Xi^{* 0} \Xi^0}.
\end{eqnarray}
They are a consequence of the underlying unitary symmetry,
and are valid for all values of the strange quark mass. 
Eq.(\ref{rel3}) can be used 
to predict the elusive decuplet couplings from the 
experimentally better known octet and decuplet-octet
transition couplings. As far as we know, these relations are new.

In summary, we have used the Morpurgo formalism to predict pion-baryon 
coupling constants. The inclusion of two-body operators leads to
significant corrections of the additive quark model values. 
Finally, we hope that this work will stimulate further research along these 
lines, such as the inclusion of three-quark 
operators and a more rigorous treatment of SU(3) flavor breaking.

\vspace{0.5 cm} 
\noindent
{\bf Acknowledgement:} 
This work has been partially supported by a U.S. DOE grant.
We would like to thank Drs. G. Morpurgo and G. Dillon for useful
criticism and valuable suggestions.

\begin{table}[t]
\label{tab1}
\begin {tabular}{|l|c|c|} \hline
Baryon & First order & Second order\\ \hline
p & $\frac{5}{3}A_1$ & -$\frac{2}{3} A_2$ \\
$\Sigma^+ $&$ \frac{4}{3} A_1$ &$\frac{2 (2-r)}{3}A_2$\\ 
$\Sigma^0 \rightarrow \Lambda^0$ &$  -\frac{2\sqrt{3}}{3} A_1$&
$\frac{2\sqrt{3}}{3}A_2$\\ 
$\Xi^0$ &$ -\frac{1}{3} A_1$ &$  \frac{4 \, r}{3} A_2$ \\ \hline 
$\Delta^+  \rightarrow p$ & $ \frac {4 \sqrt{2}}{3} A_1$ & 
- $ \frac {4 \sqrt{2}}{3}A_2$\\ 
$\Sigma^{* +} \rightarrow \Sigma^+$&$ \frac{2\sqrt{2}}{3} A_1$ 
&$\frac{2\sqrt{2}(1-2r)}{3}A_2$\\ 
$\Sigma^{* 0} \rightarrow \Lambda^0$ &$ \frac{2\sqrt{6}}{3} A_1$&
$-\frac{2\sqrt{6}}{3}A_2$\\ 
$\Xi^{* 0} \rightarrow \Xi^{0}$ 
&$ \frac{2\sqrt{2}}{3} A_1$ &$  -\frac{2\sqrt{2}\, r}{3} A_2$ \\ \hline 
$\Delta^{+}$ & $    A_1$ & $ 2 A_2$\\ 
$\Sigma^{*+}$ & $ 2 A_1$ & $ 2(1+r) A_2$\\ 
$\Xi^{*0}$ & $  A_1$ & $ 2r A_2$\\  \hline
\end{tabular}
\caption[Table 1]{Table of quark model matrix elements 
of the operator in Eq.(\ref{quarkop}) 
to first order (${\mathcal O}_1$) and second order 
corrections (${\mathcal O}_2$).}
\end{table}

\begin{table}[h]
\label{tab2}
\begin {tabular}{|l|r|r|r|} \hline
Baryon & First order  & Total  & Total \\ 
       &  ($A_2=0$)  &    r=1       &     r=0.6 \\ \hline
p & 1 & 1 & 1\\ 
$\Sigma^+ $& 0.80 &0.59    & 0.54\\
$\Sigma^0 \rightarrow \Lambda^0$ & -0.69& -0.82 & -0.82 \\
$\Xi^0$ &-0.20  &-0.42  & -0.32 \\ \hline
$\Delta^+ \rightarrow p $ & 1.70 & 2* & 2*\\
$\Sigma^{* +} \rightarrow \Sigma^+ $& 0.98 &1.16    & 0.92\\
$\Sigma^{* 0} \rightarrow \Lambda^0$ & -1.20& -1.42 & -1.42 \\
$\Xi^{* 0} \rightarrow \Xi^0$ &-1.20  &-1.42  & -1.28 \\ \hline
$\Delta^{+}$ & 0.80 & 0.23 & 0.23\\
$\Sigma^{*+}$ & 0.80 & 0.23 & 0.32\\ 
$\Xi^{*0}$    & 0.80 & 0.23 & 0.42\\ \hline 
\end{tabular}
\caption[Table 2]{Coupling constants of the pion to 
various members of the
octet and decuplet, and the decuplet-octet transitions
in terms of $f=f_{\pi^0 p}$. The * indicates an input.
The ratio $r=m_u/m_s$ of non-strange and strange quark masses
indicates the degree of flavor symmetry breaking. }
\end{table}

\begin{table}[h]
\label{tab3}
\begin{tabular} {|l|l|l|l|l|}\hline 
$ \;$&  SR \cite{Sto} & \multicolumn{2}{c|}{KDOL \cite{Kim}} & 
this work\\ \hline
 & & SU(3) & broken SU(3) & $\;$ \\ \hline 
$\Sigma^+$& 0.71 & 0.28 & 0.83 & 0.54\\
$\Sigma^0 \rightarrow \Lambda^0$ &-0.75 &--- &--- & -0.82\\
$\Xi^0$ &-0.29 & -0.46 & -2.05 &-0.32 \\ \hline 
\end{tabular}
\caption[table 3]{Coupling constants of the pion to baryons 
in terms of $f_{\pi^0 p}$ as found by various authors.  }
\end{table}

\end {document}